\documentclass[a4paper]{article}
\usepackage{amssymb}
\newcommand{\bea}{\begin{eqnarray}}
\newcommand{\eea}{\end{eqnarray}}
\newcommand{\ba}{\begin{array}}
\newcommand{\ea}{\end{array}}
\newcommand{\bc}{\begin{center}}
\newcommand{\ec}{\end{center}}
\newcommand{\be}{\begin{equation}}
\newcommand{\ee}{\end{equation}}

\newcommand{\dsf}{\displaystyle\frac}
\def\s{\sigma}

\def\Q{\mathbb{Q}}
\def\Z{\mathbb{Z}}

\def\O{\Omega}

\def\m{\mu}

\def\g{\gamma}
\def\G{\Gamma}
\begin{document}
\Large
\begin{center}
{\bf ON INHOMOGENEOUS $p$-ADIC POTTS MODEL ON A CAYLEY TREE}\\[1.1cm]
\large
{FARRUH MUKHAMEDOV}\\[1mm]
\small {\it
Department of Mechanics and Mathematics, \\
National University of Uzbekistan,\\
Vuzgorodok, 700095,  Tashkent, Uzbekistan\\
E-Mail: far75m@yandex.ru}\\[4mm]
\large \
{UTKIR ROZIKOV}\\[1mm]
\small {\it Institute of Mathematics, 29,  F.Hodjaev str.,
Tashkent, 700143,
Uzbekistan\\
E-mail: rozikovu@yandex.ru}\\[1.3cm]
\end{center}
\begin{abstract}

We consider a nearest-neighbor inhomogeneous $p$-adic Potts (with
$q\geq 2$ spin values) model on the Cayley tree of order $k\geq
1$. The inhomogeneity means that the interaction $J_{xy}$
couplings depend on nearest-neighbors points $x, y $ of the Cayley
tree. We study ($p-$ adic) Gibbs measures of the model. We show
that (i) if $q\notin p\mathbb{N}$ then there is  unique Gibbs
measure for any $k\geq 1$ and $\forall J_{xy}$ with
$|J_{xy}|<p^{-1/(p-1)}$. (ii) For $q\in p\mathbb{N}, \ \ p\geq 3$
one can choose $J_{xy}$ and $k\geq 1$ such that there exist at
least two Gibbs measures which are translation-invariant.\\[2mm]


\end{abstract}

\large

\section{Introduction}

In the paper we consider models with a nearest neighbor
interactions in the field of $p-$adic numbers on Cayley tree. The
classical (real value) example of such model is the Ising model,
with two values of spin $\pm 1$, it was considered in \cite{GR},
\cite{KT},\cite{M},\cite{MR1}.

The $p$-adic numbers were introduced by K. Hensel. Many
applications of these numbers in theoretical physics have been
proposed in papers (see for example,
\cite{ADFV},\cite{FW},\cite{K1},\cite{MP},\cite{VVZ}). A number of
$p$-adic models in physics cannot be described using ordinary
probability theory based on the Kolmogorov axioms \cite{Ko}.
$p$-adic probability models were investigated in
\cite{K2},\cite{K3}. This is a non-Kolmogorovean model in which
probabilities take values in the field of $p$-adic numbers. This
model appears to provide the probabilistic interpretation of
$p$-adic valued wave functions and string amplitudes in the
framework of $p$-adic theoretical physics (see
\cite{K1},\cite{V1}).

In \cite{KL1},\cite{KL2} the theory of stochastic processes with
values in $p$-adic and more general non-Archimedean fields having
probability distributions with non-Archimedean values has been
developed. The non-Archimedean analogue of the
Kolmogorov theorem was proved, that gives the opportunity to construct wide
classes of stochastic processes by using finite dimensional
probability distributions.

Since the probability theory and stochastic processes in a
non-Archimedean setting has been introduced, it is natural to
begin the study and the development of the problems of statistical
mechanics in the context of the $p$-adic theory of probability.

One of the central problems in the theory of Gibbs measures is to
describe infinite-volume Gibbs measures corresponding to a given
Hamiltonian. However, complete analysis of the set of Gibbs
measures for a specific Hamiltonian is often a difficult problem.
Note, that if for a given Hamiltonian there exist at least two Gibbs
measures then a {\it phase transition} is said to occur for
this model.

In this paper we will develop the $p$-adic probability theory
approaches to study inhomogeneous Potts models on a Cayley tree
over the field of $p$-adic numbers. Note in \cite{GMR} the
homogeneous $p$-adic Potts model with $q$ spin variables on the
set of integers $\mathbb{Z}$ has been considered. The aim of this
paper is to investigate Gibbs measures and a phase transition
problem for the model under consideration.

\section{The formal background}

\subsection{$p$-adic numbers }

Let $\Q$ be the field of rational numbers. Throughout the paper
$p$ will be a fixed prime number. Every rational number $x\neq 0$
can be represented in the form $x=p^r\dsf{n}{m}$, where
$r,n\in\mathbb{Z}$, $m$ is a positive integer, $(p,n)=1$,
$(p,m)=1$. The $p$-adic norm of $x$ is given by
$$
|x|_p=\left\{ \ba{ll}
p^{-r} & \ \textrm{ for $x\neq 0$}\\
0 &\ \textrm{ for $x=0$}.\\
\ea \right.
$$
It satisfies the following properties:

1) $|x|_p\geq 0$ and $|x|_p=0$ if and only if $x=0$,

2) $|xy|_p=|x|_p|y|_p$,

3) the strong triangle inequality
$$
|x+y|_p\leq\max\{|x|_p,|y|_p\},
$$
this is a non-Archimedean .

The completion of $\Q$ with  respect to the $p$-adic norm would be
a field and it is called {\it $p$-adic field} which is denoted by
$\Q_p$.

The well-known Ostrovsky's theorem asserts that norms
$|x|_{\infty}=|x|$ and $|x|_p$, $p=2,3,5...$ exhaust all
nonequivalent norms on $\Q$ (see \cite{Kl}). Any $p$-adic number
$x\neq 0$ can be uniquely represented in the canonical series: $$
x=p^{\g(x)}(x_0+x_1p+x_2p^2+...) , \eqno(2.1) $$ where
$\g=\g(x)\in\Z$ and $x_j$ are integers, $0\leq x_j\leq p-1$,
$x_0>0$, $j=0,1,2,...$ (see more detail \cite{Kl},\cite{VVZ}). In
this case $|x|_p=p^{-\g(x)}$.

Let $B(a,r)=\{x\in \Q_p : |x-a|_p< r\}$, where $a\in \Q_p$, $r>0$.
The $p$-adic logarithm is defined by series
$$
\log_p(x)=\log_p(1+(x-1))=\sum_{n=1}^{\infty}(-1)^{n+1}\dsf{(x-1)^n}{n},
$$
which converges for every $x\in B(1,1)$. And $p$-adic exponential
is defined by
$$
\exp_p(x)=\sum_{n=1}^{\infty}\dsf{x^n}{n!},
$$
which converges for every $x\in B(0,p^{-1/(p-1)})$.

{\bf Lemma 2.1.} \cite{Kl},\cite{VVZ} {\it Let $x\in
B(0,p^{-1/(p-1)})$ then we have $$ |\exp_p(x)|_p=1,\ \ \
|\exp_p(x)-1|_p=|x|_p<1, \ \ |\log_p(1+x)|_p=|x|_p<p^{-1/(p-1)} $$
and $$ \log_p(\exp_p(x))=x, \ \ \exp_p(\log_p(1+x))=1+x. $$ }

Let $(X,{\cal B})$ be a measurable space, where ${\cal B}$ is an
algebra of subsets $X$. A function $\m:{\cal B}\to \Q_p$ is said
to be a $p$-adic measure if for any $A_1,...,A_n\subset{\cal B}$
such that $A_i\cap A_j=\emptyset$ ($i\neq j$) the equality holds
$$
\mu(\bigcup_{j=1}^{n} A_j)=\sum_{j=1}^{n}\mu(A_j).
$$

A $p$-adic measure is called a probability measure if $\mu(X)=1$.
A $p$-adic probability measure $\m$ is called bounded if
$\sup\{|\m(A)|_p : A\in {\cal B}\}<\infty $.

For more detail information about $p$-adic measures we refer to
\cite{K2},\cite{K3}.

\subsection{The Cayley tree}

The Cayley tree  $\Gamma^k$ of order $ k\geq 1 $ is an infinite
tree, i.e., a graph without cycles, from each vertex of which
exactly $ k+1 $ edges issue. Let $\Gamma^k=(V, L,i)$ , where $V$
is the set of vertexes of $ \Gamma^k$, $L$ is the set of edges of
$ \Gamma^k$ and $i$ is the incidence function associating each
edge $l\in L$ with its  endpoints $x,y\in V$. If $i(l)=\{x,y\}$,
then $x$ and $y$ are called {\it neighboring vertices's}, and we
write $l=<x,y>$. A collection of the pairs
$<x,x_1>,...,<x_{d-1},y>$ is called {\sl path} from the point $x$
to the point $y$. The distance $d(x,y), x,y\in V$, on the Cayley
tree, is the length of the shortest path from $x$ to $y$.

We set
$$ W_n=\{x\in V| d(x,x^0)=n\}, $$
$$ V_n=\cup_{m=1}^n W_m=\{x\in V| d(x,x^0)\leq n\}, $$
$$ L_n=\{l=<x,y>\in L | x,y\in V_n\}, $$
for an arbitrary point $ x^0 \in V $. Denote $|x|=d(x,x_0)$, $x\in
V$.

Denote
$$
S(x)=\{y\in W_{n+1} :  d(x,y)=1 \} \ \ x\in W_n,
$$
this set is called {\it direct successors} of $x$. Observe that
any vertex $x\neq x^0$ has $k$ direct successors and $x^0$ has
$k+1$.

\indent{\bf Proposition 2.2}. \cite{G} {\it There exists a
one-to-one correspondence between the set  $V$ of vertices of the
Cayley tree of order $k\geq 1$ and the group $G_{k}$ of the free
products of $k+1$ cyclic  groups  of the second order with
generators
$a_1,a_2,...,a_{k+1}$.}\\

Let us define a group structure on the group $\G_{k}$ as follows.
Vertices which correspond to the "words" $g,h\in G_{k}$ are
called nearest neighbors and are connected by an edge if either
$g=ha_i$ or $h=ga_j$ for some $i$ or $j$. The graph thus defined
is a Cayley tree of order $k$.

Consider a left (resp. right) transformation shift on $G_{k}$
defined as: for $ g_0\in G_{k}$ we set
$$
T_{g_0}h=g_0h \ \ (\textrm{resp.}\ \  T_{g_0}h=hg_0,) \ \  \forall
h\in G_{k}.
$$
It is easy to see that the set of all left (resp. right)
shifts on $G_{k}$ is isomorphic to the group $G_{k}$.

\subsection{The inhomogeneous $p$-adic Potts model}

Let $\Q_p$ be the field of $p$-adic numbers. By $\Q_p^{q-1}$ we
denote $\underbrace{\Q_p\times...\times\Q_p}_{q-1}$.
 A norm $\|x\|_p$
of an element $x\in \Q_p^{q-1}$ is defined by
$\|x\|_p=\max\limits_{1\leq i\leq q-1}\{|x_i|_p\}$, here
$x=(x_1,...,x_{q-1})$. By $xy$ we understand a bilinear form on
$\Q_p^{q-1}$ defined by
$$
xy=\sum_{i=1}^{q-1}x_iy_i, \ \ x=(x_1,\cdots,x_{q-1}),
y=(y_1,\cdots,y_{q-1}).
$$

Let $\Psi=\{\s_1,\s_2,...,\s_q\}$, where $\s_1,\s_2,...,\s_q$ are
elements of $\Q_p^{q-1}$ such that $\|\s_i\|_p=1$,
$i=\overline{1,q}$ and
$$
\s_i\s_j= \left\{ \ba{ll}
1, \ \  \textrm{for $i=j$},\\
0, \ \ \textrm{for $i\neq j$}\\
\ea \right. (i,j=\overline{1,q-1}), \ \s_q=\sum_{i=1}^{q-1}\s_i.
$$

Let $h\in \Q_p^{q-1}$, then we have $h=\sum_{i=1}^{q-1}h_i\s_i$
and
$$
h\s_i= \left\{ \ba{ll}
h_i, \ \ \textrm{for $i=\overline{1,q-1}$},\\
\sum_{i=1}^{q-1}h_i, \ \ \textrm{for $i=q$}\\
\ea \right. \eqno(2.2)
$$

We consider $p$-adic Potts model where spin takes values in the
set $\Psi$ and is assigned to the vertices of the tree. Denote
$\O_{n}=\Psi^{V_n}$, it is the configuration space on $V_n$. The
Hamiltonian $H_n:\O_{n}\to\Q_p$ of inhomogeneous $p$-adic Potts
model has the form
$$
H_n(\s)=-\sum_{<x,y>\in L_n}J_{xy}\delta_{\s(x),\s(y)},  \ \
n\in\mathbb{N}, \eqno(2.3)
$$
here $\s=\{\s(x) : x\in V_n\}\in\O_n$, $\delta$ is the Kronecker
symbol and
$$|J_{xy}|_p<p^{-1/(p-1)}, \ \ \forall <x,y>. \eqno (2.4)$$

We say (2.3) is {\it homogeneous Potts model} if $J_{xy}=J, \ \
\forall <x,y>.$

\section{A construction of Gibbs measures}

In this subsection we give a construction of a special class of
Gibbs measures for $p$-adic Potts models on the Cayley tree.

To define Gibbs measure we need in the following

{\bf Lemma 3.1.} {\it Let $h_x, x\in V$ be a $\Q_p^{q-1}$-valued
function such that $\|h_x\|_p\leq p^{-1/(p-1)}$ for all $x\in V$
and $J_{xy}\in B(0,p^{-1/(p-1)}), \ \ <x,y>\in L_n$. Then the
relation
$$
H_n(\s)+\sum_{x\in W_n}h_x\s(x)\in B(0,p^{-1/(p-1)})
$$
is valid for any $n\in \mathbb{N}$.}

The proof easily follows from the strong triangle inequality for
the norm $\|\cdot\|_p$.

Let $h:x\in V\to h_x\in\Q_p^{q-1}$ be a function of $x\in V$ such
that $\|h_x\|_p<p^{-1/(p-1)}$ for all $x\in V$. Given $n=1,2,...$
consider a $p$-adic probability measure $\m^{(n)}_h$ on
$\Psi^{V_n}$ defined by
$$
\mu^{(n)}_h(\s_n)=Z^{-1}_{n}\exp_p\{-H_n(\s_n)+\sum_{x\in
W_n}h_x\s(x)\}, \eqno(3.1)
$$
Here, as before, $\s_n:x\in V_n\to\s_n(x)$ and $Z_n$ is the
corresponding partition function:
$$
Z_n=\sum_{\tilde\s_n\in\Omega_{V_n}}\exp_p\{-H_n(\tilde\s_n)+
\sum_{x\in W_n}h_x\tilde\s(x)\}.
$$

Note that according to Lemma 3.1 the measures $\m^{(n)}$ exist and
here the condition (2.4) is necessary to define $\exp_p$.

The compatibility condition for $\m^{(n)}_h(\s_n), n\geq 1$ is
given by the equality\footnote{The compatibility condition gives
us a possibility to construct a measure $\mu$ on whole $\Psi^V$ by
means of the measures $\m^{(n)}$ defined on $\Psi^{V_n}$ such that
the restriction of the measure $\mu$ to $\Psi^{V_n}$ coincides
with the measure $\m^{(n)}$.}
$$
\sum_{\s^{(n)}}\m^{(n)}_h(\s_{n-1},\s^{(n)})=\m^{(n-1)}_h(\s_{n-1}),
\eqno(3.2)
$$
where $\s^{(n)}=\{\s(x), x\in W_n\}$ (cp.
\cite{BRZ1},\cite{BRZ2}).

We note that an analogue of the Kolmogorov extension theorem for
distributions can be proved for $p$-adic distributions given by
(3.1) (see \cite{KL2}). If (3.2) holds for some function $h=\{h_x
: x\in V\}$ then according to the Kolmogorov theorem there exists
a unique $p$-adic measure $\m_h$ depending on $h$ and defined on
$\O=\Psi^V$ such that for every $n=1,2,...$ and
$\s_n\in\Psi^{V_n}$ the equality holds
$$
\m_h\bigg(\{\s|_{V_n}=\s_n\}\bigg)=\m^{(n)}_h(\s_n).
$$
This $\m_h$ measure is said to be {\it $p$-adic Gibbs measure} for
the considered Potts model. By ${\cal S}$ we denote the set of all
$p$-adic Gibbs measures associated with functions $h=\{h_x,\ x\in
V\}$. If $|{\cal S}|\geq 2$, then we say that for this model there
exists {\it a phase transition}, otherwise, we say there is {\it
no phase transition} ( here $|A|$ means the cardinality of a set
$A$). Now our problem is to find for what kind of functions
$h=\{h_x : x\in V\}$ the measures defined by (3.1) would satisfy
the compatibility condition (3.2). Of course, there are many
functions $h_x$ for which the condition (3.2) is not satisfied.
For example, let the order of the tree be 2, i.e. $\G^2=\Z$, and
consider homogeneous $p$-adic Potts model with $q=2$. In this case
the model reduces to the well-known Ising model with spin values
$\pm 1$. Now for an arbitrary non-zero $h\in\Q_p$, $|h|_p\leq
p^{-1/(p-1)}$ set
$$
h_n= \left\{
\begin{array}{ll}
h, \ \ \ \textrm{if} \ \ \ n=2k,\\[2mm]
0, \ \ \ \textrm{if} \ \ \ n=2k+1.\\
\end{array}
\right.
$$

Then it is not hard to verify that  the corresponding  measures
(see (3.1)) associated with the function $h=\{h_n, \ n\in\Z\}$ do
not satisfy the condition (3.2).\\

The following statement describes conditions on $h_x$ guaranteeing
the compatibility condition for the measures $\m^{(n)}_h(\s_n)$.

{\bf Theorem 3.2.} {\it The measures $\m^{(n)}_h(\s_n), \
n=1,2,...$ satisfy the compatibility condition (3.2) if and only
if for any $x\in V$ the following equation holds:
$$
h_x'=\sum_{y\in S(x)}F(h_y';\theta_{xy},q) \eqno(3.3)
$$
here and below $\theta_{xy}=\exp_p(J_{xy})$, a vector
$h'=(h_1',...,h_{q-1}')$ is defined by a vector
$h=(h_1,...,h_{q-1})$ as follows $h_i'=\sum\limits_{j=1,j\neq
i}^{q-1}h_j$, $i=1,...,q-1$ and a mapping $F:\Q_p^{q-1}\to\Q_p^{q-1}$ is \\
$F(h;\theta,q)=(F_1(h;\theta,q),...,F_{q-1}(h;\theta,q))$ with
$$
F_i(h;\theta,q)=F_i(h_1,...,h_{q-1};\theta,q)=
\log_p\bigg[\frac{(\theta-1)\exp_p(h_i)+\sum_{j=1}^{q-1}\exp_p(h_j)+1}
{\sum_{j=1}^{q-1}\exp_p(h_j)+\theta}\bigg],
$$
where $i=1,...,q-1,$ and $\theta\in \Q_p$. }

{\bf Proof.} Using (2.2) it is easy to see that (3.2) and (3.3)
are equivalent.(cf. \cite{MR1},\cite{MR2}).\\[3mm]

Observe that according to this Theorem the problem of describing
of $p$-adic Gibbs measures reduces to the describing of solutions
of functional equation (3.3).\\[3mm]

\section{Uniqueness of the Gibbs measure}

In this section we will prove the following

{\bf Theorem 4.1.} {\it If $q\notin p\mathbf{N}$ then the equation
(3.3) has unique solution $h_x=(0,...,0)\in \mathbb{Q}^{q-1}_p, \
\ \forall x\in V,$ for every $k\geq 1$ and $J_{xy}$ with (2.4),
i.e. $|{\cal S}|=1$.}

In order to prove this Theorem we will prove some auxiliary lemmas.

The following lemma plays the key role in our analysis.

{\bf Lemma 4.2.} {\it If $|a_i-1|_p\leq M$ and $|a_i|_p=1$,
$i=1,...,n$, then
$$ \bigg|\prod_{i=1}^{n}a_i-1\bigg|_p\leq
M. \eqno(4.1) $$ }

{\bf Proof.} We prove by induction on $n$. The case $n=1$ is the
condition of lemma. Suppose that (4.1) is valid at $n=m$. Now let
$n=m+1$. Then we have \bea
\bigg|\prod_{i=1}^{m+1}a_i-1\bigg|_p=\bigg|\prod_{i=1}^{m+1}a_i-
\prod_{i=1}^ma_i+\prod_{i=1}^ma_i-1\bigg|_p\leq\nonumber\\
\leq\max\bigg\{\bigg|\prod_{i=1}^na_i(a_{n+1}-1)\bigg|_p,
\bigg|\prod_{i=1}^na_i-1\bigg|_p\bigg\}\leq M\nonumber \eea

This completes the proof.

Let $h_x=(h_{1,x},\cdots ,h_{q-1,x})$, $ x\in V$ be a solution of
(3.3). Denote $z_x=(z_{1,x},...,z_{q-1,x})$, where
$z_{i,x}=\exp_p(h'_{i,x})$ or
$z_{i,x}=\exp_p\bigg(\sum\limits_{j=1,j\neq
i}^{q-1}h_{j,x}\bigg)$, $i=1,\cdots,q-1$. Then the equation (3.3)
can be written as follows
$$z_{i,x}=\prod_{y\in S(x)}{(\theta_{xy}-1)z_{i,y}+\sum^{q-1}_{j=1}z_{j,y}+1
\over \sum^{q-1}_{j=1}z_{j,y}+\theta_{xy}}, \ \ i=1,...,q-1 \eqno
(4.2)$$

Let $S(x)=\{x_1,...,x_k\}$, here as before $S(x)$ is the set of
direct successors of $x$. Using this notation we rewrite the
equation (4.2) as
$$
z_{i,x}=\prod_{m=1}^ka_{i,x,m} ,
$$
where $a_{i,x,m}= \dsf{(\theta_{xx_m}-1)
z_{i,x_m}+\sum^{q-1}_{j=1}z_{j,x_m}+1}
{\sum^{q-1}_{j=1}z_{j,x_m}+\theta_{xx_m}}$,
 $\theta_{xx_m}=\exp_p(J_{xx_m})$, $m=1,...,k$,  $x\in V$.\\

{\bf Lemma 4.3.} {\it For every $x\in V$ the following inequality
holds
$$ |h'_{i,x}|_p\leq \dsf{1}{p}\max_{1\leq j\leq
k}\{|h'_{i,x_j}|_p\}. $$}

{\bf Proof.} For every $m\in\{1,2,...,k\}$ and $i=1,...,q-1$ we
have $$|a_{i,x,m}-1|_p=
\bigg|\frac{(\theta_{xx_m}-1)(z_{i,x_m}-1)}
{\sum^{q-1}_{j=1}z_{j,x_m}+\theta_{xx_m}}\bigg|_p =
$$
$$
=\bigg|\frac{(\theta_{xx_m}-1)(z_{i,x_m}-1)}
{\sum^{q-1}_{j=1}(z_{j,x_m}-1)+(\theta_{xx_m}-1)+q}\bigg|_p\leq
\dsf{1}{p}|h'_{i,x_m}|_p.  $$ Also
$$|a_{i,x,m}|_p=
\bigg|\dsf{(\theta_{xx_m}-1)
z_{i,x_m}+\sum^{q-1}_{j=1}(z_{j,x_m}-1)+q}
{\sum^{q-1}_{j=1}(z_{j,x_m}-1)+(\theta_{xx_m}-1)+q}\bigg|_p=1.
$$ Here and above we have used the equality
$|\exp_p(x)-1|_p=|x|_p$ (see Lemma 2.1), the condition (2.4) and
$|q|_p=1$.

Hence the conditions of Lemma 4.2 are satisfied for $a_{i,x,m},
m=1,...,k$, whence
$$ |h'_{i,x}|_p=|z_{i,x}-1|_p=\bigg|\prod^k_{m=1}a_{i,x,m}-1\bigg|
\leq \dsf{1}{p}\max_{1\leq j\leq k}\{|h'_{i,x_j}|_p\}, $$ this
completes the proof.

{\bf Lemma 4.4.} {\it If $h'_x=0$ then $h_x=0$.}

The proof follows from the equality
$$
h_{k,x}=\dsf{1}{q-2}\sum_{i=1}^{q-1}h'_{i,x}-h'_{k,x},\eqno(4.3)
$$
where $k=1,\cdots,q-1$.

{\bf Proof of Theorem 4.1.} It is easy to see that $h_x=(0,...,0),
x\in V$ is a solution of (3.3). We want to prove that any other
solution of (3.3) coincides with this one.

Now let $h_x=(h_{1,x},\cdots.h_{q-1,x}), x\in V$ be a solution of
(3.3). Take an arbitrary $\varepsilon>0$. Let $n_0\in\mathbb{N}$
be such that $\dsf{1}{p^{n_0}}<\varepsilon$. According to Lemma
4.3 we have \bea \|h'_x\|_p\leq\dsf{1}{p}\|h'_{x_{i_0}}\|_p\leq
\dsf{1}{p^2}\|h_{x_{i_0,i_1}}\|_p\leq\cdots\nonumber\\
\leq\dsf{1}{p^{n_0-1}}\|h_{x_{i_0,...,i_{n_0-2}}}\|_p
\leq\dsf{1}{p^{n_0}}<\varepsilon,\nonumber \eea here
$x_{i_0,...,i_n,j}$, $j=1,..,k$ are direct successors of
$x_{i_0,...,i_n}$ and
$$\|h_{x_{i_0,...,i_m}}\|_p=\max\limits_{1\leq j\leq
k}\{\|h_{x_{i_0,...,i_{m-1},j}}\|_p\}.$$

The arbitrariness of $\varepsilon$ implies that $h'_x=0$, hence
Lemma 4.4. yields $h_x=0$ for every $x\in V$. The theorem is
proved.\\

\section{Non-uniqueness of the Gibbs measure}

In this section, we shall prove non-uniqueness of Gibbs measure for
$q\in p\mathbb{N}.$ Denote
$$
\Lambda=\{ h=(h_x\in \Q_p^{q-1}, x\in V) : h_x \ \
\textrm{satisfies the equation (3.3)}\}.
$$

To prove that the Gibbs measure is not unique it is enough to show
that set $\Lambda $ contains at least two distinct elements. The
description of an arbitrary elements of the set $\Lambda$ is a
complicated problem.

In this section, we restrict ourselves to description of periodic
elements of $\Lambda$.

Let $G_k$ be a free product of $k+1$ cyclic groups of order two.
According to Proposition 2.2 there is a one-to-one correspondence
between the set of vertices $V$ of the Cayley tree $\G^k$ and the
group $G_k$. Let $\hat G_k\subset G_k$ be a normal subgroup of
finite index.

We say that $h=\{h_x : x\in G_k\}$ is {\it $\hat G_k$-periodic} if
$h_{yx}=h_x$ for all $x\in G_k$ and $y\in\hat G_k$. A Gibbs
measure is called {\it $\hat G_k$-periodic} if it corresponds to
$\hat G_k$-periodic function $h$. Note that $G_k$-periodic Gibbs
measures are called {\it translation-invariant}.

Let $H_0$ be a subgroup of index $r$ in $G_k$, and let
$G_k|H_0=\{H_0,H_1,...,H_{r-1}\}$ be the quotient group. Let
$q_i(x)=|S^*(x)\cap H_i|, i=0,1,...,r-1$; $N(x)=|\{j:q_j(x)\ne
0\}|,$ where $S^*(x)$ is the set of all nearest neighbors of $x\in
G_k.$ Denote $Q(x)=(q_0(x),q_1(x),...,q_{r-1}(x)).$ We note (see
\cite{R}) that for every $x\in G_2$ there is a permutation $\pi_x$
of the coordinates of the vector $Q(e)$ (where $e$ is the identity
of $G_k$) such that
$$ \pi_xQ(e)=Q(x). $$
It follows from this equality that $N(x)=N(e)$ for all $x\in G_k.$

Each $H_0-$ periodic function is given by
$$\{h_x=h^{(i)} \ \ \mbox{for} \ \ x\in H_i, i=0,1,...,r-1\}.$$

Let $G^*_k$ be the subgroup in $G_k$ consisting of all words of
even length. Clearly, $G^*_k$ is a subgroup of index 2. We are
interested in the answer to the following question: whether there exist at
least two $G^*_k$-periodic $p$-adic Gibbs measures?

For the simplicity we will consider the cases $k=1,2$ with $q\in p\mathbb{N}$.\\

{\tt Case: $k=1.$} In this case, we assume
$$ \theta_{xy}=
\left\{ \ba{ll} \theta_1  \ \ \textrm{if} \ \  x \in G^*_1, \ \
y\in G_1
\setminus G_1^*,\\[2mm]
\theta_2  \ \ \textrm{if} \ \ x\in G_1\setminus G^*_1, \ \
 y \in G^*_1
\ea \right. \eqno(5.1) $$ and
$$ h_x=
\left\{ \ba{ll} h=(h_1,...,h_{q-1})  \ \
\textrm{if} \ \  x\in G^*_1, \\[2mm]
l=(l_1,...,l_{q-1}) \ \ \textrm{if} \ \ x\in G_1\setminus G^*_1,
 \ea \right.$$

Then from (3.3) we have
$$ \left\{ \ba{ll}
z_i=\dsf{(\theta_1-1) t_i+\sum^{q-1}_{j=1}t_j+1}{\sum^{q-1}_{j=1}t_j+\theta_1} \\[2mm]
t_i=\dsf{(\theta_2-1) z_i+\sum^{q-1}_{j=1}z_j+1}{\sum^{q-1}_{j=1}z_j+\theta_2} \\[2mm]
\ea \right. \eqno(5.2) $$
where $z_i=\exp_p(h'_i)$ and $t_i=\exp_p(l'_i).$ Observe that if
$z_i=1$ then $t_i=1$ (the converse is also true).  Denote
$\alpha={\theta_1\theta_2+q-1\over \theta_1+\theta_2+q-2}$, then
substituting $t_i=1, \ \ i=2,...,q-1$ into (5.2) we get
$$z={\alpha z+q-1\over z+\alpha+q-2},\eqno (5.3)$$
with $z=z_1$.

It is easy to see that the equation (5.3) has two solutions $z=1$
and $z=1-q$ for any $\alpha\ne 1$ and $z=1$ if $\alpha=1.$

Note that a solution of (5.3) will define a $p$-adic Gibbs measure
if it satisfies the inequality $|z-1|_p\leq {1\over p},$ which is
true for $z=1$ and $z=1-q$, since $q\in p\mathbb{N}.$ Thus the
system of equations (5.2) has two solutions $z_i=1$, $t_i=1$,
$i=1,\cdots,q-1$ and
$z_1=1-q,z_j=1$, $t_1=1-q,t_j=1$, $j=2,\cdots, q-1$ if $\alpha\ne 1$.
Consequently, the equation (3.3) has two solutions (see (4.3)).\\[4mm]

{\bf Remark 5.1.} It is known \cite{Ru} that for the Potts model
and even for arbitrary models on $\mathbb{Z}$ with finite radius
of interaction of the particles, in which $\mathbb{R}$ is
considered instead of $\Q_p$, there are no phase transitions. In
the case under consideration this pattern is destroyed.

{\tt Case: $k=2.$} In this case for simplicity we will assume
that $\theta_{xy}=\theta$ for any $<x,y>.$ Then the problem of
describing $G^*_2$-periodic Gibbs measures reduces to the
description of solutions of the following equation:
$$
\left\{ \ba{ll} h^{(1)'}=F(h^{(2)'};\theta,q)\\[2mm]
h^{(2)'}=F(h^{(1)'};\theta,q),\\
\ea \right. \eqno(5.4)
$$
where $h^{(1)'},h^{(2)'}\in\Q_p^{q-1}$ and the map $F$ is defined in
Theorem 3.2.

Observe that for every $l=1,...,q-1$ $h^{(i)'}_l=0$ satisfies
$l$-th equation, where
$h^{(i)'}=(h^{(i)'}_1,\cdots,h^{(i)'}_{q-1})$, $i=1,2$. Denoting
$z_i=\exp_p(h^{(i)'}_1)$ and substituting $h^{(i)'}_l=0$ at
$l=2,...,q-1$ to the first equation of (5.4) we obtain
$$
\left\{ \ba{ll} z_1=f(z_2)\\[2mm]
z_2=f(z_1),\\
\ea \right. \eqno(5.5)
$$
here
$$
f(z)=\biggl(\frac{\theta z+q-1}{z+\theta+q-2}\biggr)^2. \eqno(5.6)
$$

The first equation in (5.5) can be written as follows:
$$(\theta^2-z_1)z_2^2+[2\theta(q-1)-2(\theta+q-2)z_1]z_2
-(\theta+q-2)^2z_1+(q-1)^2=0. \eqno (5.7)$$

We know that if $z_2=z_1$ then (5.7) reduces to the equation
$$z_1^3+(2q-(\theta-1)^2-3)z_1^2+((\theta-1)^2+3+q^2-4q)z_1
-(q-1)^2=0, \eqno (5.8)$$ which describes the
translation-invariant Gibbs measures.

We add and subtract (5.8) in (5.7), consequently, we have
$$(z_2-z_1)\{(\theta^2-z_1)z_2-z^2_1+(\theta^2-2\theta+4-2q)z_1+
2\theta (q-1)\}-$$
$$-[z_1^3+(2q-(\theta-1)^2-3)z_1^2+((\theta-1)^2+3+q^2-4q)z_1
-(q-1)^2]=0. \eqno (5.9)$$

Now substituting $z_2=f(z_1)$ in (5.9) and noticing that the
numerator of $z_2-z_1=f(z_1)-z_1$ equal to the sentence in the
square brackets, so cutting them out we get
$$(\theta^2-z_1)f(z_1)-z^2_1+(\theta^2-2\theta+4-2q)z_1+2\theta(q-1)+
(z_1+\theta+q-2)^2=0. $$ or
$$(-\theta^2+z_1)f(z_1)=\theta^2 z_1+(\theta^2+q^2+4\theta
q-6\theta-4q+4). $$ Hence we get
$$
(\theta^2+\theta+q-2)^2z_1^2+$$
$$[\theta^4+4(q-1)\theta^3+(q^2+6q-12)\theta^2+2(5q^2-18q+16)\theta
+(2q^3-13q^2+26q-17)]z_1+$$
$$+[\theta(q-1)+(\theta+q-2)^2]^2=0.
\eqno(5.10)$$ Denote
$$
\alpha=(\theta^2+\theta+q-2)^2,
$$
$$
\beta=\theta^4+4(q-1)\theta^3+(q^2+6q-12)\theta^2+2(5q^2-18q+16)\theta
+(2q^3-13q^2+26q-17),
$$
$$
\tau=[\theta(q-1)+(\theta+q-2)^2]^2.
$$
The following equalities
$$
\alpha=[(\theta^2-1)+(\theta-1)+q]^2,
$$
$$
\beta=(\theta^4-1)+4(q-1)(\theta^3-1)+(q^2+6q-12)(\theta^2-1)+
2(5q^2-18q+16)(\theta-1)+2q^3-2q^2,
$$
$$
\tau=[3(q-1)(\theta-1)+(\theta-1)^2+(q-1)^2]^2
$$
and the inequality $|\theta^i-1|_p\leq\frac{1}{p}$,
$i\in\mathbb{N}$ imply with the strong triangle property that
$$
|\alpha|_p\leq\dsf{1}{p}, \ \ \ |\beta|_p\leq \dsf{1}{p}, \ \ \
|\tau|_p=1, \eqno(5.11)
$$
here we have used that $q\in p\mathbb{N}$.

A solution $z$ of (5.10) will define a $p$-adic Gibbs measure if
it satisfies the inequality $|z-1|_p\leq \frac{1}{p}$. So, we
should check the last condition. Rewrite the equation (5.10) as
follows
$$
\alpha(z_1^2-1)+\beta(z_1-1)+\alpha+\beta+\tau=0.\eqno(5.12)
$$
The equalities (5.11) imply that
$$
|\alpha(z_1^2-1)+\beta(z_1-1)|\leq \dsf{1}{p}, \ \
|\alpha+\beta+\tau|_p=1
$$
this means, that the equation (5.12) has no solutions, which
satisfy $|z-1|_p\leq \frac{1}{p}$.

Now, for the simplicity we restrict ourselves to the case: $q=3$ and
$p=3$. In this case, it is easy to see that the equation (5.4) can
have solutions only of the form $(h,0)$, $(0,h)$ and $(h,h)$.
 According to above
made argument, one can see that the equation (5.4) has no solution
like $(h,0)$ and $(0,h)$. Solutions of the form $(h,h)$ describes
only translation-invariant Gibbs measures.

Summarizing, we obtain the following

{\bf Theorem 5.1.} {\it (i) For $k=1$ and $q\in pN$ any $G^*_1-$
periodic Gibbs measures of inhomogeneous $p$-adic Potts model with
condition (5.1) coincide with translation-invariant Gibbs
measures. If $(\theta_1-1)(\theta_2-1)\neq 0$ then there occurs a
phase transition.

(ii) If $k=2$, $p=3$, $q=3$ then for 3-adic homogeneous Potts
model  $G_2^*$-periodic Gibbs measures coincide with
translation-invariant ones which correspond to the
solutions of the equation (5.8), in this case there is a phase transition.}\\

{\bf Remark 5.2.} In the real case it is known \cite{MR1} that the
$G_2^*$-periodic Gibbs measures are different from
translation-invariant ones even for the homogeneous  Ising model
on the Cayley tree $\Gamma^2$.\\

Note, that existence of translation-invariant Gibbs measures for
homogeneous $p$-adic Potts models has been proved in \cite{GMR}
for $k=1$ and in \cite{MR2} for $k=2$. The main results of these
papers are the following\\

{\bf Theorem 5.2.} \cite{GMR} {\it If $k=1$ and $q\in p\mathbb{N},
\ \ p>2$ then for $p$-adic Potts model there are (phase
transitions) at least two translation-invariant
Gibbs measures.}

{\bf Theorem 5.3.} \cite{MR2} (i) {\it Let $p=2$, $q\in
2^2\mathbb{N}$ and $J\neq 0$. If $q=2^2s$, $(s,2)=1$,
$|J|_2=\dsf{1}{4}$ or $q=2^ms$, $m\geq 3$, $(s,2)=1$,
$|J|_2\leq\dsf{1}{4}$ then there exists (phase transition) at
least two translation-invariant Gibbs measures for the homogeneous
$2$-adic Potts model on a Cayley tree of order 2.}

(ii) {\it Let $p\geq 3$, $q\in p\mathbb{N}$, and
$0<|J|_p\leq\dsf{1}{p}$ then there exist at least $q$
translation-invariant Gibbs  measures for the homogeneous $p$-adic
Potts model on a Cayley tree of order 2.}

(iii) {\it The $p$-adic Gibbs measure corresponding to the
homogeneous $p$-adic Potts model on the Cayley tree of order $k$
is bounded if and only if $q\notin p\mathbb{N}$.}\\

{\bf Conjecture.} For any $k\geq 1$, $q\in p\mathbb{N}$ and any
subgroup of finite index, each periodic Gibbs measure of the
homogeneous $p$-adic Potts model is translation-invariant.\\

{\bf Remark 5.3.} If $q\notin p\mathbb{N}$ then Theorem 5.3 says
that the $p$-adic Gibbs measure corresponding to the Potts model
is bounded, hence  in this case there is (unique)
bounded $p$-adic Gibbs measure  for the $p$-adic Potts model.\\

{\bf Remark 5.4.} If $q=2$ then $p$-adic Potts model becomes the
$p$-adic Ising model. Hence, Theorems 4.1 and 5.3 imply for $p\geq
3$ that for the inhomogeneous $p$-adic Ising model there is unique
Gibbs measure which is translation-invariant and bounded. The
description of Gibbs measures for the inhomogeneous $p$-adic Ising
model with $p=2$ would be given elsewhere.\\

{\bf Acknowledgment} The work was done within the scheme of Borsa
di Studio NATO-CNR. The first named author (F.M.) thanks the
Italian CNR for providing financial support and II Universita di
Roma "Tor Vergata" for all facilities. The second named author
(U.R.) thanks Institute des Hautes Etudes Scientifiques (IHES) for
supporting the visit to IHES, Bures-sur-Yvette in
September-December 2003 and the IGS programme at the Isaac Newton
Institute (INI) for supporting the visit to INI in November 2003.
The authors also acknowledge their gratitude to Professors I.V.
Volovich and A.Yu. Khrennikov for the helpful comments and
discussions.

The authors are grateful to the referee's helpful suggestions.


\begin{thebibliography}{99}

\bibitem{ADFV} I.Ya.Aref'eva, B.Dragovich, P.H.Frampton, I.V.Volovich, - Wave
function of the universe and $p$-adic gravity, Int. J. Mod.Phys. A
{\bf 6} 4341-4358 (1991).

\bibitem{BRZ1} P.M. Bleher, J. Ruiz and V.A. Zagrebnov, - On the purity of the limiting
Gibbs state for the Ising model on the Bethe lattice, Jour.
Statist. Phys. {\bf 79}: 473-482 (1995).

\bibitem{BRZ2} P.M. Bleher, J. Ruiz and  V.A. Zagrebnov, - On the phase diagram of the
random field Ising model on the Bethe lattice, Jour. Statist.
Phys. {\bf 93}: 33-78 (1998).

\bibitem{G} N.N. Ganikhodjaev, Group representations and
automorphisms of the Cayley tree, Dokl. Akad. Nauk. Rep.
Uzbekistan, {\bf 4}: 3-5 (1994).

\bibitem{GMR} N.N.Ganikhodjaev, F.M.Mukhamedov, U.A.Rozikov, -
Existence of a phase transition for the Potts $p$-adic model on
the set $\mathbb{Z}$, Theor.Math.Phys. {\bf 130} 425-431 (2002).

\bibitem{GR} N.N. Ganikhodjaev, U.A. Rozikov,  On disordered phase in
the ferromagnetic Potts model on the Bethe lattice,  Osaka Jour.
Math. {\bf 37} 373-383 (2000).

\bibitem{FW} P.G.O.Freund, E.Witten, - Adelic string ampletudes,
Phys. Lett. {\bf B199}, 191-194 (1987).

\bibitem{KT} S.Katsura, M. Takizawa, -  Bethe lattice and Bethe
approximation. Progr. Theor. Phys. {\bf 51}  82-98 (1974).

\bibitem{K1} A.Yu.Khrennikov, -  $p$-adic quantum mechanics with $p$-adic valued
functions, J.Math.Phys. {\bf 32} 932-936 (1991).

\bibitem{K2} A.Yu.Khrennikov, - $p$-adic Valued Distributions in Mathematical Physics
(Kluwer Academic Publisher, Dordrecht, 1994).

\bibitem{K3} A.Yu.Khrennikov,  - $p$-adic valued probability measures, Indag. Mathem. N.S.
{\bf 7} 311-330 (1996).

\bibitem{KL1} A.Yu.Khrennikov, S.Ludkovsky - On infinite products of
non-Archimedean measure spaces. Indag. Mathem. N.S. {\bf 13}
177-183 (2002).

\bibitem{KL2} A.Yu.Khrennikov, S.Ludkovsky  Stochastic process on
non-Archimedean spaces with values in non-Archimedean fields. Adv.
Stud. in Contemp. Math. {\bf 5:1}, 57-91 (2002).

\bibitem{Kl} N.Koblitz, - $p$-adic numbers, $p$-adic analysis and
zeta-function (Berlin: Springer, 1977).

\bibitem{Ko} A.N.Kolmogorov, -  Foundation of the Theory of Probability (Chelsea, 1956).

\bibitem{MP} E.Marinary, G.Parisi, -  On the $p$-adic five point function, Phys.Lett.
{\bf 203B} 52-56 (1988).

\bibitem{M} M. Miyamoto,  - Spitzer's Markov chains with measurable
potentials. J. Math. Kyoto Univ. {\bf 22} 41-69 (1982).

\bibitem{MR1} F.M. Mukhamedov, U.A. Rozikov, - On Gibbs measures of models
with competing ternary and binary interactions and corresponding
von Neumann algebras. J. Stat. Phys. {\bf 114} 825-848 (2004).

\bibitem{MR2} F.M. Mukhamedov, U.A. Rozikov,- On Gibbs measures of $p-$ adic
Potts model on the Cayley tree. Indag. Mathem. N.S. (To appear).

\bibitem{R} U.A. Rozikov, Partition structures of the group representation of the
Cayley tree into cosets by finite-index normal subgroups and their
applications to the description of periodic Gibbs distributions,
{\it Theor. Math. Phys.} {\bf 112}: 929-933 (1997).

\bibitem{Ru} D.Ruelle, -  Statistical Mechanics: Rigorus Results
(Benjamin, 1969).

\bibitem{VVZ} V.S.Vladimirov, I.V.Volovich, E.I.Zelenov, -  $p$-adic Analysis and
Mathematical Physics. (Singapour: World Scientific, 1994).

\bibitem{V1} I.V.Volovich, -  $p$-adic strings, Class. Quantum Grav. {\bf 4},
L83-L87 (1987).



\end{thebibliography}
\end{document}